# Towards an Active, Autonomous and Intelligent Cyber Defense of Military Systems: the NATO AICA Reference Architecture

Paul Theron
Thales
Salon de Provence, France
paul.theron@thalesgroup.com

Alexander Kott
U.S. Army Research Laboratory
Adelphi, MD, USA
alexander.kott1.civ@mail.mil

Martin Drašar
Masaryk University
Brno, Czech Republic
drasar@ics.muni.cz

Krzysztof Rzadca
University of Warsaw
Warsaw, Poland
krzadca@mimuw.edu.pl

Benoît LeBlanc
Ecole Nationale Supérieure de Cognitique
Bordeaux, France
benoit.leblanc@ensc.fr

Mauno Pihelgas
NATO CCDCOE
Tallinn, Estonia
mauno.pihelgas@ccdcoe.org

Luigi Mancini
Sapienza University
Rome, Italy
mancini@di.uniroma1.it

Agostino Panico
Sapienza University
Rome, Italy
panico@di.uniroma1.it

*Abstract*—Within the future Global Information Grid, complex massively interconnected systems, isolated defense vehicles, sensors and effectors, and infrastructures and systems demanding extremely low failure rates, to which human security operators cannot have an easy access and cannot deliver fast enough reactions to cyber-attacks, need an active, autonomous and intelligent cyber defense. Multi Agent Systems for Cyber Defense may provide an answer to this requirement. This paper presents the concept and architecture of an Autonomous Intelligent Cyber defense Agent (AICA). First, we describe the rationale of the AICA concept. Secondly, we explain the methodology and purpose that drive the definition of the AICA Reference Architecture (AICARA) by NATO's IST-152 Research and Technology Group. Thirdly, we review some of the main features and challenges of Multi Autonomous Intelligent Cyber defense Agent (MAICA). Fourthly, we depict the initially assumed AICA Reference Architecture. Then we present one of our preliminary research issues, assumptions and ideas. Finally, we present the future lines of research that will help develop and test the AICA / MAICA concept.

I. RATIONALE FOR THE AICA/MAICA CONCEPT

Today, five broad types of systems coexist in Land, Sea and Air operations:
- Office and information management systems, which includes web services, emailing systems, and information management applications ranging from human resource management to logistics through maintenance and project management;
- C4ISR systems for the command of war operations, with associated Battlefield Management Systems that extend the C4ISR down to single vehicles and platoons;
- Communication systems such as SATCOM, L16, line of sight networks, software defined radios, etc.;
- Platform and life automation systems, similar to industrial systems and that provide sea vessels for instance with air conditioning, refrigeration, lifts, video surveillance, etc.;
- Weapon systems, which include both sensors and effectors of all kinds, including the Internet of Battle Things (IoBT).

On the battlefield, the future Global Information Grid will mix those technologies into complex large scale networks of massively interconnected systems, the cybersecurity supervision of which by human operators will become increasingly difficult, if not impossible.

Besides, a great number of military missions will require defense vehicles and effectors to work stealthily while some will find themselves isolated because of poor bandwidth or because communications will become untrustworthy. Isolated systems will create a specific class of problems in terms of the possibility to monitor and manage their cybersecurity. On one hand, to fully analyze their cyber-health would be possible only when connected at base during maintenance and operation preparation. On the other hand, in case of cyber-attacks, they will require immediate counter-reactions while no cybersecurity or cyber defense specialist is available.

Finally, defense infrastructures and systems engaged in battle operations must show extremely low failure rates. Counter reactions to cyber-attacks must therefore be initiated at the speed of operation of these systems, not at the (low) speed of human decision making in the presence of complex issues. In a conflict with a technically sophisticated adversary, military tactical networks will be a heavily contested battlefield. Enemy software cyber agents -- malware -- will infiltrate friendly networks and attack friendly C4ISR and computerized weapon systems.

In this context, systems' cyber defense will be organized in two manners:

- Connected systems of lesser criticality will be monitored by cybersecurity sensors, security information and event management (SIEM) systems, and security operations centers (SOCs). This will be the case of office and information management systems and of C4ISR systems under peaceful circumstances.
- Higher grade systems or configurations such as C4ISR systems deployed in combat circumstances, communication systems, life and automation systems and weapon systems require autonomous intelligent cyber defense capabilities.

To fight cyber-attacks that may target this last class of military systems, we expect that NATO needs artificial cyber hunters - intelligent, autonomous, mobile agents specialized in active cyber defense, that we call Autonomous Intelligent Cyber defense Agents (AICA). They will work in cohorts or swarms and will be capable, together, to detect cyber-attacks, devise the appropriate counter measures, and run and adapt tactically their execution.

Those friendly NATO cyber agents -- goodware -- will stealthily patrol networks, detect enemy agents while remaining concealed, and they will devise the appropriate counter-attack plan and then destroy or degrade the enemy malware. They will do so mostly autonomously, because human cyber experts will be always scarce on the battlefield, because human reactions will be too slow, and because connectivity might be nonexistent or poor.

Agents will be learning and adaptive because the enemy malware and attack patterns are constantly evolving. They will be stealthy because the enemy malware will try to find and kill them. They will work in cohorts or swarms as attacks will be sophisticated and stealthy, and only collective intelligence will stand a chance to detect the early combined signs of malware actions and positions. In addition, they will do so because combatting malware will mean fighting a variety of pieces of malware acting either simultaneously or in a sequence hard to detect, and intelligently spread across the friendly military systems and networks they attack to produce the effect sought by the enemy.

Deployed on NATO networks, the AICA friendly software agents will become a major force multiplier. The agents will augment the inevitably limited capabilities of human cyber defenders, and will team with humans when ordered or in need to do so. Without such agents, the effective defense of NATO computer networks and systems might become impossible if attackers also resort on multi agent systems to carry out their attacks. Without active autonomous intelligent cyber defense agents, a NATO C4ISR will not survive an encounter with a determined, technically sophisticated enemy.

At this time, such capabilities remain unavailable for the defensive purposes of NATO. To acquire and successfully deploy such agents, in an inter-operable manner, NATO Nations must create a common technical vision - reference architecture - and a roadmap.

## II. PURPOSE AND METHODOLOGY OF THE PROJECT

Inspired by the above rationale, NATO's IST-152 Research and Technology Group (RTG) is an activity that was initiated by the NATO Science and Technology Organization and was kicked-off in September 2016. The group is developing a comprehensive, use case focused technical analysis methodology in order to produce a first-ever reference architecture and technical roadmap for active autonomous intelligent cyber defense agents. In addition, the RTG is working to identify and evaluate selected elements that may be eligible contributors to such capabilities and that begin to appear in academic and industrial research.

Scientists and engineers from several NATO Nations have brought unique expertise to this project. Only by combining multiple areas of distinct expertise along with a realistic and comprehensive approach can such a complex software agent be provided.

The output of the RTG may become a tangible starting point for acquisition activities by NATO Nations. If based on a common reference architecture, software agents developed or purchased by different Nations will be far more likely to be interoperable.

## III. MAIN FEATURES AND CHALLENGES OF THE MAICA CONCEPT

Related research includes Mayhem (from DARPA Cyber Challenge, but also Xandra, etc.), agents from the Pechoucek's group, Professor Mancini's work on the AHEAD architecture [1] and the Aerospace Cyber Resilience research chair's research program [2], Anti-Virus tools (Kaspersky, Bitdefender, Avast, Norton, etc. etc.), HBSS, OSSEC, Various host-based IDS/IPS systems, Application Performance Monitoring Agents, Anti-DDOS systems and Hypervisors. Also, a number of related research directions include topics such as deep learning (especially if it can be computationally inexpensive), Botnet technology (seen as a network of agents), network defense games, flip-it games, the Blockchain, and fragmentation and replication. The introduction of Artificial Intelligence into military systems, such as C4ISR, has been studied, for instance by [3] and [4]. Multi Agent Systems form an important part of AI.

Since the emergence of the concept of Multi Agent Systems (e.g., [5]), MAS have been deployed in a number of contexts such as power engineering [6] and their decentralized automated surveillance [7], industrial systems [8], networked and intelligent embedded systems [9], collective robotics [10], wireless communication [11], traffic simulation and logistics planning [12], home automation [13].

However, if the use of intelligent agents for the cyber defense of network-centric environments has already long been envisaged [14], effective research in this area is still new.

In the context of the cyber defense of friendly systems, an "agent" has been defined [2] as a piece of software or hardware, an autonomous processing unit:

- With an individual mission and the corresponding competencies, i.e. in analyzing the milieu in which the agent is inserted, detecting attacks, planning the required countermeasures, or steering and adapting tactically the execution of the latter, or providing support to other agents like for instance inter-agent communication;
- With proactivity, i.e. the capacity to engage into actions and campaigns without the need to be triggered by another program or by a human operator;
- With autonomy, i.e. a decision making capacity of its own, the capacity to function or to monitor, control and repair itself on its own, without the need to be controlled by another program or by a human operator, and the capacity to evaluate the quality of its own work and to adjust its algorithms in case of deviance from its norm or when its rewards (satisfaction of its goals) get poor;
- Driven by goals, decision making and other rules, knowledge and functions fit for its purpose and operating circumstances;
- Learning from experience to increase the accuracy of its decisions and the power of its reactions;
- With memories (input, process, output, storage);

- With perception, sensing and action, and actuating interfaces;
- Built around the adequate architecture and appropriate technologies;
- Positioned around or within a friendly system to defend, or patrolling across a network;
- Sociable, i.e. with the capacity to establish contact and to collaborate with other agents, or to enter into a cyber cognitive cooperation when the agent requires human help or to cooperate with a central Cyber C2;
- Trustworthy, i.e. that will not deceive other agents nor human operators;
- Reliable; i.e. that do what they are meant to do, during the time specified and under the conditions and circumstances of their concept of operation;
- Resilient, i.e. both robust to threats (including cyber-threats aimed at disabling or destroying the agent itself; the agent being able to repel or withstand everyday adverse events and to avoid degrading), and resistant to incidents and attacks that may hit and affect the agent when its robustness is insufficient (i.e. the agent is capable of recovering from such incidents or attacks);
- Safe, i.e., conceived to avoid harming the friendly systems the agent defends, for instance by calling upon a human supervisor or central cyber C2 to avoid making wrong decisions or to adjust their operating mode to challenging circumstances, or by relocating when the agent is the target of an attack and if relocation is feasible and allows protecting it, or by activating a fail-safe mode, or by way of self-destruction when no other possibility is available.

In the same context (ibid), a multi agent system is a set of agents:

- Distributed across the parts of the friendly system to defend;
- Organized in a swarm (horizontal coordination) or cohort (vertical coordination);
- In which agents may have homogeneous or heterogeneous roles and features;
- Interoperable and interacting asynchronously in various ways such as indifference, cooperation, competition;
- Pursuing a collective non-trivial cyber defense mission, i.e. allowing to piece together local elements of situation awareness or propositions of decision, or to split a counter-attack plan into local actions to be driven by individual agents;
- Capable of self-organization, i.e. as required by changes in circumstances, whether external (the attack's progress or changes in the friendly system's health or configuration) or internal (changes in the agents' health or status);
- That may display emergent behaviors [15], i.e. performances that are not explicitly expressed in individual agents' goals, missions and rules; in the context of cyber defense, "emergence" is likely to be an interesting feature as, consisting in the "*way to obtain dynamic results, from cooperation, that cannot easily be predicted in a deterministic way*" [15]; it can be disturbing to enemy software in future malware-goodware "tactical" combats within defense and other complex systems;
- Extensible or not, i.e. open or closed to admitting new agents in the swarm or cohort;
- Safe, trustworthy, reliable and resilient as a whole, which is a necessity in the context of cyber defense whereas in other, less challenging contexts may be unnecessary. Resilience, here, may require maintaining a system of virtual roles as described in a human context by [16].

AICA will not be simple agents. Their missions, competencies, functions and technology will be a challenging construction in many ways.

Among many such challenges, we can mention [2] working in constrained environments, the design of agents' architecture and the attribution of roles and possible specialization to each of them, agents' decision making process [17], the capacity to generate and execute autonomously plans of counter-measures in case of an attack, agents' autonomy, including versus trustworthiness, MAICA's safety to defense systems, cyber cognitive cooperation [18], agents' resilience in the face of attacks directed at them by enemy software, agents' learning capacities and the development of their functional autonomy, the specification and emergence of collective rules for the detection and resolution of cyber-attacks, AICA agents' deployment concepts and rationale, their integration into host hardware as [8] showed in industrial system contexts, etc.

IV. THE INITIAL AICA REFERENCE ARCHITECTURE

To start the research with an initial assumption about agents' architecture, the IST-152-RTG designed the AICA Reference Architecture on the basis of classical perspective reflected in [19] and [20].

At the present moment, it is assumed to include the following functional components:

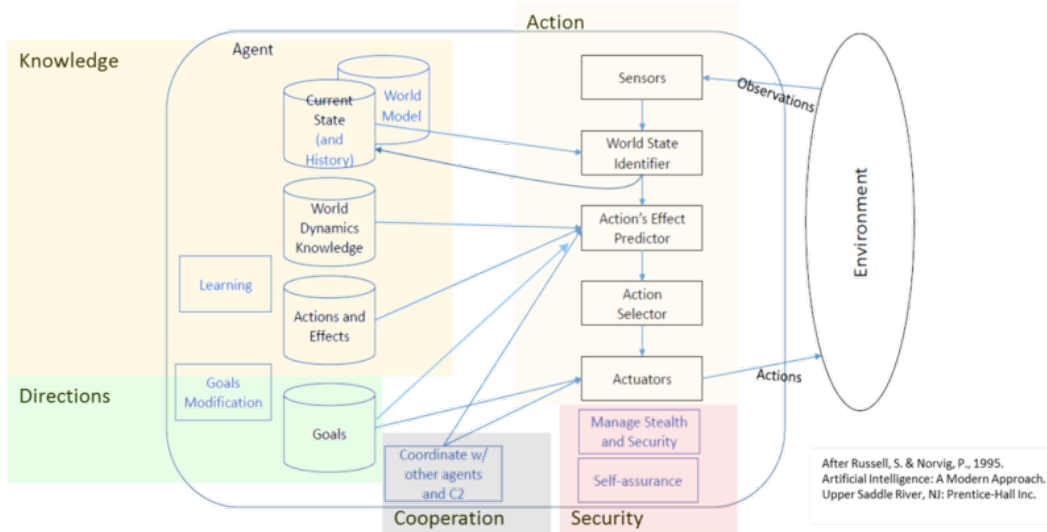

Figure 1 Assumed functional architecture of the AICA

The AICA Reference Architecture delivers five main high-level functions:

- Sensing and world state identification.
- Planning and action selection.
- Collaboration and negotiation.
- Action execution.
- Learning and knowledge improvement.

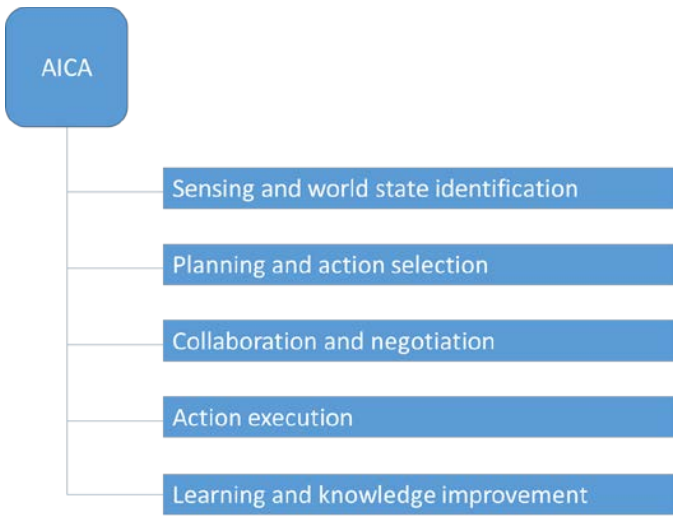

Figure 2 The AICA's main five high-level functions

*A. Sensing and World state identification*

DEFINITION: Sensing and World state identification is the AICA's high-level function that allows a cyber-defense agent to acquire data from the environment and systems in which it operates as well as from itself in order to reach an understanding of the current state of the world and, should it detect risks in it, to trigger the Planning and Action selection high-level function. This high-level function relies upon the "World model", "Current state and history", "Sensors" and "World State Identifier" components of the assumed functional architecture.

The Sensing and World state identification high-level function includes two functions: (1) Sensing; (2) Word state identification.

a-1. Sensing

DESCRIPTION: Sensing operates from two types of data sources: (1) External (system and device-related) current world state descriptors; (2) Internal (agent-related) current state descriptors.

Current world state descriptors, both external and internal, are captured on the fly by the agent's Sensing function. They may be double-checked, formatted or normalized for later use by the World state identification function (to create processed current state descriptors).

a-2. World state identification

DESCRIPTION: The World state identification function operates from two sources of data: (1) Processed current state descriptors; (2) Learnt world state patterns.

Learnt world state patterns are stored in the agent's world knowledge repository. Processed current state descriptors and Learnt world state patterns are compared to identify problematic current world state patterns (i.e. presenting an anomaly or a risk). When identifying a problematic current world state pattern, the World state identification function triggers the Planning and Action selection high-level function.

b. Planning and action selection

DEFINITION: Planning and action selection is the AICA's high-level function that allows a cyber-defense agent to elaborate one to several action proposals and to propose them to the Action selector function that decides the action or set of actions to execute in order to resolve the problematic world state pattern previously identified by the World state identifier function. This high-level function relies upon the "World dynamics", "Actions and effects", "Goals", "Actions' effect predictor" and "Action selector" components of the assumed functional architecture.

The Planning and action selector high-level function includes two functions: (1) Planning; (2) Action selector.

b-1. Planning

DESCRIPTION: The Planning function operates on the basis of two data sources: (1) Problematic current world state pattern; (2) Repertoire of actions (Response repertoire).

The Problematic current world state pattern and Repertoire of actions (Response repertoire) are concurrently explored in order to determine the action or set of actions (Proposed response plan) that can resolve the submitted problematic current world state pattern. The action or set of actions so determined are presented to the Action selector. It may be possible that the Planning function requires some form of cooperation with human operators (cyber cognitive cooperation, C3).

It may alternatively require cooperation with other agents or with a central cyber C2 (command and control) in order to come up with an optimal set of actions forming a global response strategy. Such cooperation could be either to request from other agents or from the cyber C2 complementary action proposals, or to delegate to the cyber C2 the responsibility of coordinating a global set of actions forming the wider response strategy.

It may be possible that the Planning function requires some form of cooperation with human operators (cyber cognitive cooperation, C3). It may alternatively require cooperation with other agents or with a central cyber C2 (command and control) in order to come up with an optimal set of actions forming a global response strategy. Such cooperation could be either to request from other agents or from the cyber C2 complementary action proposals, or to delegate to the cyber C2 the responsibility of coordinating a global set of actions forming the wider response strategy.

These aspects have been the object of an initial study in [17] where options such as offline machine learning, pattern recognition, online machine learning, escalation to a human operator, game theoretic option search, and failsafe have been envisaged, and in [18] for cyber cognitive cooperation processes.

b-2. Action selector

DESCRIPTION: The Action selector function operates on the basis of three data sources: (1) Proposed response plans; (2) Agent's goals; (3) Execution constraints and requirements, e.g., the environment's technical configuration, etc.

The proposed response plan is analyzed by the Action selector function in the light of the agent's current goals and of the execution constraints and requirements that may either be part of the world state descriptors gained through the Sensing and World state identifier high-level function or be stored in the agent's data repository and originated in the Learning and Knowledge improvement high-level function. The proposed response plan is then trimmed from whatever element does not fit the situation at hand, and augmented of prerequisite, preparatory or precautionary or post-execution recommended complementary actions. The Action selector thus produces an Executable Response Plan, and then submitted to the Action execution high-level function.

Like with the Planning function, it is possible that the Action selector function requires to liaise with human operators, other agents or a central cyber C2 (command and control) in order to come up with an optimal Executable Response Plan forming part of and being in line with a global response strategy. Such cooperation could be to exchange and consolidate information in order to come to a collective agreement on the assignment of the various parts of the global Executable Response Plan and the execution responsibilities to specific agents. It could alternatively be to delegate to the cyber C2 the responsibility of elaborating a consolidated Executable Response Plan and then to assign to specific agents the responsibility of executing part(s) of this overall plan within their dedicated perimeter. This aspect is not yet studied in the present release of the AICA Reference Architecture.

c. Collaboration and negotiation

DEFINITION: Collaboration and negotiation is the AICA's high-level function that allows a cyber-defense agent 1) to exchange information (elaborated data) with other agents or with a central cyber C2, for instance when one of the agent's functions is not capable on its own to reach satisfactory conclusions or usable results, and 2) to negotiate with its partners the elaboration of a consolidated conclusion or result. This high-level function relies upon the "Coordinate with other agents and C2" component of the assumed functional architecture.

The Collaboration and negotiation high-level function includes, at the present stage, one function: Collaboration and negotiation.

DESCRIPTION: The Collaboration and negotiation function operates on the basis of three data sources: (1) Internal, outgoing data sets (i.e. sent to other agents or to a central C2); (2) External, incoming data sets (i.e. received from other gents or from a central cyber C2); (3) Agents' own knowledge (i.e. consolidated through the Learning and knowledge improvement high-level function).

When an agent's Planning and action selector function or other function needs it, the agent's Collaboration and negotiation function is activated. Ad hoc data are sent to (selected) agents or to a central C2. The receiver(s) may be able, or not, to elaborate further on the basis of the data received through their own Collaboration and negotiation function. At this stage, when each agent (including possibly a central cyber C2) has elaborated further conclusions, it should share them with other (selected) agents, including (or possibly not) the one that placed the original request for collaboration. Once this (these multiple) response(s) received, the network of involved agents would start negotiating a consistent, satisfactory set of conclusions. Once an agreement reached, the concerned agent(s) could spark the next function within their own decision making process.

When the agent's own security is threatened the agent's Collaboration and negotiation function should help warning other agents (or a central cyber C2) of this state.

Besides, the agent's Collaboration and negotiation function may be used to receive warnings from other agents that may trigger the agent's higher state of alarm.

Finally, the agent's Collaboration and negotiation function should help agents discover other agents and establish links with them.

d. Action execution

DEFINITION: The Action execution is the AICA's high-level function that allows a cyber-defense agent to effect the Action selector function's decision about an Executable Response Plan (or the part of a global Executable Response Plan assigned to the agent), to monitor its execution and its effects, and to provide the agents with the means to adjust the execution of the plan (or possibly to dynamically adjust the plan) when and as needed. This high-level function relies upon the "Goals" and "Actuators" components of the assumed functional architecture.

The Action execution high-level function includes four functions:

- Action effector;
- Execution monitoring;
- Effects monitoring;
- Execution adjustment.

d-1. Action effector

DESCRIPTION: The Action effector function operates on the basis of two data sources:

- Executable Response Plan;
- Environment's Technical Configuration.

Taking into account the Environment's Technical Configuration, the Action effector function executes each planned action in the scheduled order.

d-2. Execution monitoring

DESCRIPTION: The Execution monitoring operates on the basis of two data sources:

- Executable Response Plan;
- Plan execution feedback.

The Execution monitoring function should be able to monitor (possibly through the Sensing function) each action's execution status (for instance: done, not done, and wrongly done). Any status apart from "done" should trigger the Execution adjustment function.

d-3. Effects monitoring

DESCRIPTION: The Effects monitoring function operates on the basis of two data sources: (1) Executable Response Plan; (2) Environment's change feedback.

It should be able to capture (possibly through the Sensing function) any modification occurring in the plan execution's environment. The associated dataset should be analyzed or explored. The result of such data exploration might provide a positive (satisfactory) or negative (unsatisfactory) environment change status. Should this status be negative, this should trigger the Execution adjustment function.

d-4. Execution adjustment

DESCRIPTION: The Execution adjustment function operates on the basis of three data sources: (1) Executable Response Plan; (2) Plan execution feedback and status; (3) Environment's change feedback and status.

The Execution adjustment function should explore the correspondence between the three data sets to find alarming associations between the implementation of the Executable Response Plan and its effects. Should warning signs be identified, the Execution adjustment function should either adapt the actions' implementation to circumstances or modify the plan.

e. Learning and knowledge improvement

DEFINITION: Learning and knowledge improvement is the AICA's high-level function that allows a cyber-defense agent to use the agent's experience to improve progressively its efficiency with regards to all other functions. This high-level function relies upon the Learning and Goals modification components of the assumed functional architecture.

The Learning and knowledge improvement high-level function includes two functions: (1) Learning; (2) Knowledge improvement.

e-1. Learning

DESCRIPTION: The Learning function operates on the basis of two data sources: (1) Feedback data from the agent's functioning; (2) Feedback data from the agent's actions.

The Learning function collects both data sets and analyzes the reward function of the agent (distance between goals and achievements) and their impact on the agent's knowledge database. Results feed the Knowledge improvement function.

e-2. Knowledge improvement

DESCRIPTION: The Knowledge improvement function operates on the basis of two data sources: (1) Results (propositions) from the Learning function; (2) Current elements of the agent's knowledge.

The Knowledge improvement function merges Results (propositions) from the Learning function and the Current elements of the agent's knowledge.

V.     PRELIMINARY RESEARCH ASSUMPTIONS AND QUESTIONS: THE EXAMPLE OF LEARNING

The environment of the agent can change rapidly, especially (but not exclusively) due to an enemy action. In addition, the enemy malware, its capabilities and Tactics, Techniques and Procedures (TTP), evolve rapidly. Therefore, the agent must be capable of autonomous learning. The reasoning capabilities of the agent rely on its knowledge bases (KBs). The purpose of the learning function(s) of the agent is to modify the KBs of the agent in a way that enhances the success of the agent's actions. The agent learns from its experiences. Therefore, the most general cycle of the learning process is the following:

1. The agent possesses a KB.
2. The agent uses the KB to perform actions; he also makes observation (receives percepts). These together constitute the agent's experience.
3. The agent uses this experience to learn the desirable modifications to the KB.
4. The agent modifies the KB.

5. Repeat.

The agent's experience needs a formal representation. It may look like this sequence:

($t_1$, $a_1$, $e_1$, $R_1$) ($t_2$, $a_2$, NULL, NULL) ($t_3$, NULL, $e_3$, $R_3$) … ($t_n$, $a_n$, $e_n$, $R_n$)

Where $t_1$ is the time when the agent starts to record his experience and $t_n$ is the moment "now", an is an action, $e_n$ is a percept, $R_n$ is the reward of the action.

To make the representation of knowledge more compact and useful, we could divide it into shorter chunks where each chunk ends with the moment when the agent is able to determine a reward. We could call such a chunk an episode. Episode $E_j$ is a sequence of pairs $\{a_1, e_i\}$, and the resulting reward $R_j$:

$E_j = (\{a_i, e_i\}, R_j)$

The following is an example of such a short episode: a1 - check file system integrity; e1 - find unexpected file; a2 - delete file; e2 - file gone; a3 – NULL; e3 - observe Enemy C2 traffic; Reward - 0.09

A representation of this nature could be used in a case-based reasoning, or in a deep learning approach.

What exactly could an agent learn? One, fairly general option is that Learning Module learns the World Dynamics model which is a function that takes as an input a state and an action applied to that state; its output is a new state that will result from application of that action, or a distribution of states. World Dynamics Model is used in particular in "Action Selector and Predictor" module. In addition, the Learning Module can learn another function required in "Action Selector and Predictor" module, which maps the current world state to a set of feasible actions.

## VI. CONCLUSIONS AND FUTURE RESEARCH

Intelligent, partly autonomous agents are likely to become primary cyber fighters on the future battlefield. Our initial exploration identified the key functions, components and their interactions for a potential reference architecture of such an agent.

The AICA Reference Architecture was derived from [19] for we needed a broad, cognition-based, all-encompassing agent structure. Future works will challenge this initial choice.

Embedding AICA agents into highly constrained military systems is also the focus of future research and this issue was not addressed at this stage yet.

And, at the present stage, the AICARA architecture is a preliminary proposal. Its feasibility as well as its power to fight malware autonomously and intelligently remain to be evaluated.

With respect to further efforts, this research group plans to have a basic proof-of-concept prototype developed and tested by 2019. The current priorities are:

- To study use cases as a reference for the research, as this will lead to clarifying the scope, concepts, functionality and functions' inputs and outputs of AICA and MAICA systems; use cases will be based on the one elaborated in the IST-152 intermediary report [21];
- To refine the initially assumed architecture by drawing further lessons from the case studies;
- To determine the set of technologies that AICAs should embark and that need to be tested during the prototyping phase;
- To define the methodology of the tests.

The sum of challenges presented by the AICA / MAICA concept appears, today, very substantial, although our initial analysis suggests that the required technical approaches do not seem to be entirely beyond the current state of the research. An empirical research program and collaboration of multiple teams should be able to produce significant results and solutions for a robust, effective intelligent agent. This might happen within a time span that could currently be assumed on the order of ten years.

REFERENCES

[1]  F. De Gaspari, S. Jajodia, L. V. Mancini and A. Panico, "AHEAD: A New Architecture for Active Defense," SafeConfig'16, October 24 2016, Vienna, Austria, 2016.

[2] P. Théron, La cyber résilience, un projet cohérent transversal à nos trois thèmes, et la problématique particulière des Systèmes Multi Agent de Cyber Défense, Leçon inaugurale, 5 décembre 2017, ed., France, Salon de Provence: Chaire Cyber Résilience Aérospatiale (Cyb'Air), 2017.

[3] R. Rasch, A. Kott and K. D. Forbus, "AI on the battlefield: An experimental exploration," *AAAI/IAAI,* 2002.

[4] R. Rasch, A. Kott and K. D. Forbus, "Incorporating AI into military decision making: an experiment," *IEEE Intelligent Systems,* vol. 18, no. 4, pp. 18-26, 2003.

[5] J. Von Neumann, "The General and Logical Theory of Automata," in *Cerebral Mechanisms in Behavior: The Hixon Symposium, September 1948, Pasadena*, L. A. Jeffress, Ed., New York, John Wiley & Sons, Inc, 1951, pp. 1-31.

[6] S. D. McArthur, E. M. Davidson, V. M. Catterson, A. L. Dimeas, N. D. Hatziargyriou, F. Ponci and T. Funabashi, "Multi-Agent Systems for Power Engineering Applications - Part I: Concepts, Approaches, and Technical Challenges," *IEEE TRANSACTIONS ON POWER SYSTEMS,* vol. 22, no. 4, pp. 1743-1752, 2007.

[7] A. Carrasco, M. C. Romero-Ternero, F. Sivianes, M. D. Hernández and J. I. Escudero, "Multi-agent and embedded system technologies applied to improve the management of power systems," *JDCTA,* vol. 4, no. 1, pp. 79-85, 2010.

[8] M. Pechoucek and V. Marík, "Industrial deployment of multi-agent technologies: review and selected case studies," *Autonomous Agents and Multi-Agent Systems,* vol. 17, p. 397–431, 2008.

[9] W. Elmenreich, "Intelligent methods for embedded systems," in *Proceedings of the First Workshop on Intelligent Solutions in Embedded Systems*, J. 2. Vienna University of Technology 2003, Ed., Austria: Vienna, Vienna University of Technology, 2003, pp. 3-11.

[10] H.-P. Huang, C.-C. Liang and C.-W. Lin, "Construction and soccer dynamics analysis for an integrated multi-agent soccer robot system," *Natl. Sci. Counc. ROC(A),* vol. 25, pp. 84-93, 2001.

[11] J.-P. Jamont, M. Occello and A. Lagrèze, "A multiagent approach to manage communication in wireless instrumentation systems," *Measurement,* vol. 43, no. 4, pp. 489-503, 2010.

[12] B. Chen and H. H. Cheng, "A review of the applications of agent technology in traffic and transportation systems," *Trans. Intell. Transport. Sys.,* vol. 11, no. 2, pp. 485-497, 2010.

[13] J.-P. Jamont and M. Occello, "A framework to simulate and support the design of distributed automation and decentralized control systems: Application to control of indoor building comfort," in *IEEE Symposium on Computational Intelligence in Control and Automation*, Paris, France, IEEE, 2011, pp. 80-87.

[14] M. R. Stytz, D. E. Lichtblau and S. B. Banks, "Toward using intelligent agents to detect, assess, and counter cyberattacks in a network-centric environment," Institute For Defense Analyses, Alexandria, VA, 2005.

[15] J.-P. Muller, "Emergence of collective behaviour and problem solving," in *Engineering Societies in the Agents World IV*, A. Omicini, P. Petta and J. Pitt, Eds., volume 3071, Lecture Notes in Computer Science, 2004, pp. 1-20.

[16] K. Weick, "The Collapse of Sensemaking in Organizations: The Mann Gulch Disaster," *Administrative Science Quarterly,* vol. 38, no. 4, pp. 628-652, 1993.

[17] B. Blakely and P. Theron, *Decision flow-based Agent Action Planning,* Prague, 2017.

[18] B. LeBlanc, P. Losiewicz and S. Hourlier, *A Program for effective and secure operations by Autonomous Agents and Human Operators in communications constrained tactical environments,* Prague, 2017.

[19] S. Russell and P. Norvig, Artificial Intelligence: A Modern Approach, Upper Saddle River, NJ: Prentice-Hall Inc, 1995.

[20] S. J. Russell and P. Norvig, Artificial Intelligence: A Modern Approach, 2nd ed. ed., Upper Saddle River, New Jersey: Prentice Hall, 2003.

[21] A. Kott, L. V. Mancini, P. Theron, M. Drašar, H. Günther, M. Kont, M. Pihelgas, B. LeBlanc and K. Rzadca, "Initial Reference Architecture of an Intelligent Autonomous Agent for Cyber Defense," US Army Research Laboratory, ARL-TR-8337, March 2018, available from https://arxiv.org/abs/1803.10664, Adelphi, MD, 2018.